\documentclass{iopart}

\begin{document}

\title{Gradient descent learning in and out of equilibrium}
\author{Nestor Caticha and Evaldo Ara\'{u}jo de Oliveira}
\address{Instituto de F\'{\i}sica - Universidade de S\~{a}o Paulo, CP 66318
S\~{a}o Paulo, SP, CEP05389-970 Brazil}

\begin{abstract}
Relations between the off thermal equilibrium dynamical process of on-line
learning and the thermally equilibrated off-line learning are studied for
potential gradient descent learning. The approach of Opper to study on-line
Bayesian algorithms is extended to potential based or maximum likelihood
learning. We look at the on-line learning algorithm that best approximates
the off-line algorithm in the sense of least Kullback-Leibler information
loss. It works by updating the weights along the gradient of an effective
potential different from the parent off-line potential. The interpretation
of this off equilibrium dynamics holds some similarities to the cavity
approach of Griniasty. We are able to analyze networks with non-smooth
transfer functions and transfer the smoothness requirement to the potential.
\end{abstract}


The application of Statistical Mechanics to the study of learning in Neural
Networks (NN) stems from the fact that the extraction of information from
data (examples) can be modeled by a dynamical process of minimization of an
energy function, possibly in the presence of (thermal) noise. In the case
where the system is allowed to equilibrate, roughly all the possible
information has been extracted from the data by the learning algorithm. In a
very important sense learning theory is different from e.g. magnetism. In
the latter the interactions are fixed by the physical constraints, and the
equilibrium state and how it is reached is the object of study. In the
former, the energy function can be chosen in order to achieve a certain
property in the equilibrium state, such as largest possible typical
generalization or memorization capability.

Techniques originated in the study of disordered systems, such as the
replica and cavity methods, TAP equations, as well as Monte Carlo
techniques, have been borrowed and extended, leading to several results in
what has become known as Off-line learning (OfL). Since disordered systems
may take too long to equilibrate, implying a high computational cost, the
search for efficient nonequilibrium learning algorithms has been undertaken.
An interesting class of methods - where essentially, examples are used one
at a time - is collected under the name of On-line learning (OnL)\cite
{amari67}. These bring the possibility of efficient performance and low
computational cost.

Recently Opper \cite{opper96}\cite{opper98} offered a new theoretical way of
studying the relation between OfL and OnL. He applied his ideas to Bayes
learning. The posterior probability distribution for the set of weights
obtained after $T$ examples is used as the prior for the next example. If
the full posterior is maintained, any calculation amounts to an OfL one. But
by projecting the posterior into a restricted family of parametric
distributions, huge computational gains can be achieved, transforming the
process into an effective OnL one. Now only a set of parameters and an
auxiliary set of hyperparameters have to be updated. The changes in the
hyperparameters induce automatically an effective annealing of tensorial
learning rates. In the case of continuous weights, he applied these ideas by
projecting to a gaussian space of posteriors. Solla and Winther \cite
{sollaw98} generalized it by extending it so that information about e.g. the
binary nature of weights can be included in a consistent way. This is simply
achieved by projecting into another family of posteriors and again imposing
that the information loss be minimized.

There is however no reason to limit these studies to the case of Bayes
learning and the aim of this paper is to extend Opper's method to include
the problem of learning by gradient descent. We obtain equations that
describe the evolution of the weights and hyperparameters for general
differentiable potentials. Then we look at some applications. We analyze the
relation between the off-equilibrium and thermal equilibrium for a special
case which is Bayes optimal with a nondifferentiable transfer function, the
noiseless Boolean perceptron, a case which cannot be treated by Opper's
Bayesian analysis. The on-line algorithm is automatically annealed and we
discuss how the annealing is related to a  performance estimate. Finally, we
apply the resulting equations to the same architecture but for a nonsmooth
potential in order to study the resulting algorithm. 

Let $y_k$ be an example. In the case of supervised learning it is to be
thought of as an input-output pair $y_k=(S_k,\sigma _k)$ and we assume that
the data pairs are generated by a map $\sigma =f_{w^{*}}(S)$ which might be
deterministic or stochastic so as to include the possibility of noise
corrupted data. For unsupervised learning or density estimation it is an
input vector $y_k=S_k$. The learning set is formed by $\mu $ such random
examples $D_\mu =(y_1,y_{2,}\ldots ,y_\mu ),$ drawn independently from
identical distributions$.$ The purpose of learning is to make an estimate $%
{\bf \hat{w}}$ of the true $N$ dimensional vector of parameters or weights $%
{\bf w}^{*}${\bf \ }. To do so a cost function or potential $V(\sigma
,f_w(S))=V\left( {\bf w},y\right) $ is introduced. Usually one seeks a
minimum of the total energy $E\left( w\right) =\sum_{k=1}^\mu V(\sigma
_k,f_w(S_k))$, so that learning is stated as an optimization problem$.$ The
additive form is adequate in the case of independent (or non interacting)
examples. There is also the possibility that aside from the learning set,
other information about the possible weight vectors is available. It might
be encoded in the prior probability $p_0({\bf w})$, that is, the probability
that can be attributed to any ${\bf w},$ of being the true parameter vector,
based on information other than $D_\mu $ . The information contained in the
prior and in the learning set can be taken into account simultaneously by
using Bayes theorem and imposing the equivalence of the minimum energy
prescription and that of maximizing the likelihood of the examples, which as
shown by Levin et al \cite{levin90} leads to a functional equation whose
solution is the Gibbs distribution : 
\begin{eqnarray}
P_V({\bf w}|D_\mu ) &=&\frac 1{Z_\mu }p_o({\bf w})P(D_\mu |{\bf w}) \\
&=&\frac 1{Z_\mu }p_o({\bf w})e^{-\beta \sum_{k=1}^\mu V({\bf w},y_k)},
\end{eqnarray}
where $\beta $ measures the sensibility of the likelihood and of course
plays the role of the inverse temperature and the partition function is
given by $Z_\mu =\int p_o({\bf w}^{\prime })P(D_\mu |{\bf w}^{\prime })d^N%
{\bf w}^{\prime }.$

The problem has been thus formulated as one of Statistical Mechanics of
disordered systems due to the random nature of the data. Spin glass behavior
for this type of system has been found in many different cases. Estimation
of parameters may turn into a computational hard problem, as suggested by
the long thermalization times encountered while doing Monte Carlo estimates.
This also happens for the prediction of the output $\sigma $ to a new
(statistically independent) input vector. A neural network, on the other
hand, once it has been trained, and a reasonable ${\bf \hat{w}}\;$been
determined, permits rapid estimation of $\sigma $. The fact that the
determination of ${\bf \hat{w},}$ using the full Gibbs distribution, may
itself be hard, seems to imply that there is no way out. However suppose a
reasonable estimate has been achieved for a learning set $D_\mu $, then the
incorporation of the information carried by a new example $y_{\mu +1}$ can
be efficiently and easily done at least in an approximate way. This is the
idea behind OnL and we now study this from the same perspective Opper has
used to analyze Bayes learning. That these estimates are in general hard to
do, leads to an approximation of the Gibbs distribution $P_V({\bf w}|D_\mu )$
by $P_g({\bf w}|D_\mu ).$ The type of problem dictates what is a useful
approximation. In many cases the fluctuations, at least for large $\mu $
will be gaussian and so we study this case. Still the approximation can be
done in many ways. To limit the loss of hard gained information, as measured
by the Kullback-Leibler \cite{kullback59} divergence, we follow \cite
{opper96}\cite{opper98}\cite{sollaw98} and project the current version of
the Gibbs distribution to a gaussian with the same mean ${\bf \hat{w}}(\mu )$
and covariance $C_{ij}(\mu ).$

OnL proceeds by storing all the information in the previous $\mu $ examples
in the vector ${\bf \hat{w}}(\mu ).$ Other auxiliary quantities, (in this
case the covariance $C_{ij}(\mu )$) usually termed hyperparameters will be
needed and their natural appearance and evolution justify naturally the
annealing of learning rates.

The basic idea is to consider the Gibbs distribution as the prior for the
new, the $\left( \mu +1\right) ^{th}$ example. Even when $P_V(w|D_\mu )$ is
substituted by the gaussian $P_g({\bf w}|D_\mu ),$ in general $P_V({\bf w}%
|D_{\mu +1})$ will not be gaussian.\ Therefore it is projected into a
gaussian of mean ${\bf \hat{w}}(\mu +1)$ and covariance $C_{ij}(\mu +1)$ The
procedure can then be iterated to include the next example. Of course this
update will change the covariance of the posterior, leading to new set of
equations relating $C_{ij}(\mu +1)$ and $C_{ij}(\mu ).$

The introduction of a new example, if the system is allowed to thermalize,
can be the starting point for a cavity analysis as studied by Griniasty \cite
{griniasty93}. We do not, by doing the gaussian approximation, allow the
system to thermalize.

In order to calculate the approximate change in the expected value of ${\bf w%
}$ , start with \cite{levin90}

\begin{equation}
P_V({\bf w}|D_{\mu +1})=\frac{P_V({\bf w}|D_\mu )e^{-\beta V({\bf w},y_{\mu
+1})}}{\int P_V({\bf w}^{\prime }|D_\mu )e^{-\beta V({\bf w}^{\prime
},y_{\mu +1})}d^N{\bf w}^{\prime }}
\end{equation}
and substitute it by 
\begin{equation}
\tilde{P}_V({\bf w}|D_{\mu +1})=\frac{P_g({\bf w}|D_\mu )e^{-\beta V({\bf w}%
,y_{\mu +1})}}{\int P_g({\bf w}^{\prime }|D_\mu )e^{-\beta V({\bf w}^{\prime
},y_{\mu +1})}d^N{\bf w}^{\prime }},
\end{equation}
then project $\tilde{P}_V({\bf w}|D_{\mu +1})$ to $P_g(w|D_{\mu +1}).$ Call
the initial conditions to this iteration procedure ${\bf \hat{w}}(0)$ for
the mean and for covariance, ${\bf C}(0)$. We call our current estimates of
the weights and the covariance ${\bf \hat{w}}(\mu )$ and ${\bf C}(\mu )$
respectively. Then
\begin{equation}
\hat{w}_i(\mu +1)=\int w_iP_g({\bf w}|D_{\mu +1})d^N{\bf w}, 
\end{equation}
\begin{equation}
C_{ij}(\mu +1)=\int \left( w_i-\hat{w}_i(\mu +1)\right) \left( w_j-\hat{w}%
_j(\mu +1)\right) P_g({\bf w}|D_{\mu +1})d^N{\bf w}^{\prime }. 
\end{equation}
Let ${\bf u}$ measure the gaussian fluctuations of ${\bf w}$ around ${\bf 
\hat{w}}(\mu )$ 
\begin{eqnarray*}
\hat{w}_i(\mu +1) &=&\frac{\int w_iP_g({\bf w}|D_\mu )e^{-\beta V(w,y_{\mu
+1})}d^N{\bf w}}{\int P_g({\bf w}^{\prime }|D_\mu )e^{-\beta V(w^{\prime
},y_{\mu +1})}d^N{\bf w}^{\prime }} \\
&=&\hat{w}_i(\mu )+\frac{\int u_ie^{-\frac 12{\bf u}^t{\bf C}^{-1}{\bf u}%
}e^{-\beta V({\bf \hat{w}}(\mu )+{\bf u},y_{\mu +1})}d^N{\bf u}}{\int
e^{-\frac 12{\bf u}^t{\bf C}^{-1}{\bf u}}e^{-\beta V({\bf \hat{w}}(\mu )+%
{\bf u},y_{\mu +1})}d^N{\bf u}}.
\end{eqnarray*}
Note that $u_ie^{-\frac 12{\bf u}^t{\bf C}^{-1}{\bf u}}=-C_{ij}\partial
_{u_j}\left( e^{-\frac 12{\bf u}^t{\bf C}^{-1}{\bf u}}\right) ,$ then one
integration by parts leads to 
\[
\hat{w}_i(\mu +1)=\hat{w}_i(\mu )+C_{ij}\frac{\int e^{-\frac 12{\bf u}^t{\bf %
C}^{-1}{\bf u}}\partial _{u_j}e^{-\beta V({\bf \hat{w}}(\mu )+{\bf u},y_{\mu
+1})}d^N{\bf u}}{\int e^{-\frac 12{\bf u}^t{\bf C}^{-1}{\bf u}}e^{-\beta V(%
{\bf \hat{w}}(\mu )+{\bf u},y_{\mu +1})}d^N{\bf u}}, 
\]
where a summation over repeated indices is implied. Note the very important
assumption that the potential is differentiable. This prevents the
application to some popular non differentiable potential based algorithms.
However we can deal with networks with a nonsmooth transfer function. Then
using 
\begin{equation}
\partial _{u_i}f\left( {\bf \hat{w}+u}\right) =\partial _{\hat{w}_i}f\left( 
{\bf \hat{w}+u}\right) ,  \label{mudadev}
\end{equation}
the on-line algorithm that results is 
\begin{equation}
\hat{w}_i(\mu +1)=\hat{w}_i(\mu )+C_{ij}(\mu )\partial _j\ln <e^{-\beta V(%
\hat{w}(\mu )+u)}>,  \label{algoritmo}
\end{equation}
where $<\cdot \cdot \cdot >$ means the average with respect to the gaussian
distribution with zero mean and covariance $C_{ij}(\mu )$ .

The next step is to determine the evolution of the covariance. In terms of
the gaussian distributed fluctuations $u_i$ of zero mean and the variation $%
\Delta \hat{w}=\hat{w}_i(\mu +1)-\hat{w}_i(\mu ),$ given by equation (\ref
{algoritmo}) 
\begin{equation}
C_{ij}(\mu +1)=\int (u_i-\Delta \hat{w}_i)(u_j-\Delta \hat{w}_j)P_g(w|D_\mu
)d^Nw^{\prime }. 
\end{equation}
Now use the identity 
\begin{equation}
u_iu_je^{-\frac 12{\bf u}^t{\bf C}^{-1}{\bf u}}=C_{ij}e^{-\frac 12{\bf u}^t%
{\bf C}^{-1}{\bf u}}+C_{ik}C_{jl}\partial _{u_k}\partial _{u_l}\left(
e^{-\frac 12{\bf u}^t{\bf C}^{-1}{\bf u}}\right), 
\end{equation}
then two integrations by parts and the use of eq. (\ref{mudadev}) determines
the prescription for the covariance update. 
\begin{equation}
C_{ij}(\mu +1)=C_{ij}(\mu )+C_{ik}(\mu )C_{jl}(\mu )\partial _k\partial
_l\ln <e^{-\beta V(\hat{w}(\mu )+u)}>.  
\label{annealing}
\end{equation}
On one hand, this set of equations describe a first (gaussian) approximation
to the problem of OfL learning with the potential $E_\mu =\sum_{\mu =1}^\mu
V\left( w;y_\mu \right) .$ On the other hand it describes an OnL learning
prescription for the update of the weight vector, and a set of
hyperparameters which are useful in improving performance.

We now consider the widely popular class of problems where the network is a
classifier into two categories $\sigma =\pm 1$ and the dimension of $S$ is $%
N $.\ We study the case where the potential $V\left( \lambda \right) $ is a
differentiable function of the stability $\lambda =\sigma {\bf w}\cdot S/%
\sqrt{N}$. How is the resulting algorithm related to the usual OnL schemes?
Let $t=\sigma {\bf \hat{w}}\cdot S/\sqrt{N}$, denote the stability of an
example previous to its presentation to the network so that $\lambda
=t+\sigma {\bf u}\cdot S/\sqrt{N},\;$the stability of example $S$ in the
network parametrized by ${\bf w}$. Introduce $\tilde{C}_{ij}(\mu )=\beta
C_{ij}(\mu )$ and ${\sf x}=S_i\tilde{C}_{ij}S_j/N$. An explicit form for $%
<\exp (-\beta V)>$ can be obtained. Introduce a $1$ in the form $1=\int
d\lambda \delta \left( \lambda -\sigma {\bf w}\cdot S/\sqrt{N}\right)
\propto \int d\lambda d\hat{\lambda}\exp i\hat{\lambda}\left( \lambda
-\sigma {\bf w}\cdot S/\sqrt{N}\right) $. A pair of quadratic integrations
show that 
\begin{equation}
<\exp (-\beta V)>\propto \int d\lambda \exp -\beta [V(\lambda )+\frac{%
(\lambda -t)^2}{2{\sf x}}], 
\end{equation}
thus for the estimate of the weights we have: 
\begin{equation}
\hat{w}_i(\mu +1)=\hat{w}_i(\mu )+\frac 1{\beta \sqrt{N}}\tilde{C}_{ij}(\mu
)S_j\sigma (\mu )\partial _t\ln \int d\lambda \exp -\beta [V(\lambda )+\frac{%
(\lambda -t)^2}{2{\sf x}}],  \label{algoritmo2}
\end{equation}
while for the annealing equation 
\begin{equation}
\tilde{C}_{ij}(\mu +1)=\tilde{C}_{ij}(\mu )+\frac 1{\beta N}\tilde{C}%
_{ik}(\mu )\tilde{C}_{jl}(\mu )S_kS_l\partial _t^2\ln \int d\lambda \exp
-\beta [V(\lambda )+\frac{(\lambda -t)^2}{2{\sf x}}].  \label{annealing2}
\end{equation}
To compare to previous work we look at the zero temperature limit. The $%
\lambda $ integral can be calculated by the saddle point method. Let $%
\lambda _o\left( t\right) \;$be the minimum of $V(\lambda )+(\lambda -t)^2/2%
{\sf x,}$ that is the solution of 
\begin{equation}
{\left[ \frac{\partial V}{\partial \lambda }+\frac{\lambda -t}{{\sf x}}%
\right] }_{_{\lambda =\lambda _o}}=0,  \label{lambda0}
\end{equation}
then $\ln <\exp (-\beta V)>=-\beta \left( V(\lambda _o)+\frac{(\lambda
_o-t)^2}{2{\sf x}}\right) .$ Define what we will show to be the effective
on-line potential 
\begin{equation}
{\cal E}_{{\sf x}}\left( t\right) \equiv V(\lambda _o)+\frac{(\lambda _o-t)^2%
}{2{\sf x}}.  \label{effective}
\end{equation}
Note that from eqs. (\ref{lambda0}) and (\ref{effective}) it is easy to see
that 
\begin{equation}
{\left. \frac{\partial V}{\partial \lambda }\right| }_{\lambda =\lambda _o}=%
\frac{\partial {\cal E}_{{\sf x}}\left( t\right) }{\partial t}.
\label{modulation1}
\end{equation}

The algorithm equations can now be written as 
\begin{equation}
\hat{w}_i(\mu +1)=\hat{w}_i(\mu )-\frac 1{\sqrt{N}}\tilde{C}_{ij}(\mu
)S_j\sigma (\mu )\frac{\partial {\cal E}_{{\sf x}}\left( t\right) }{\partial
t},  \label{deltaw}
\end{equation}
\begin{equation}
\tilde{C}_{ij}(\mu +1)=\tilde{C}_{ij}(\mu )-\frac 1N\tilde{C}_{ik}(\mu )%
\tilde{C}_{jl}(\mu )S_kS_l\frac{\partial ^2{\cal E}_{{\sf x}}\left( t\right) 
}{\partial t^2}.  \label{deltac}
\end{equation}
The update of ${\bf \hat{w}}$ (eq. \ref{deltaw}) can be identified with an
annealed (time or number of examples $\mu $ dependent $\tilde{C}_{ij}$)
tensorial learning rate Hebbian-like algorithm modulated by $\frac{\partial V%
}{\partial \lambda }|_{\lambda _o},$ the gradient of the original potential
calculated, not at the point $t\;$where it would be expected since it is the
pre-training stability, but at the posterior stability $\lambda _o.$
However, the need to calculate the gradient at a future point $\lambda _o$
would render this algorithm useless. But in its stead (see eq. (\ref
{modulation1})) the gradient $\frac{\partial {\cal E}_{{\sf x}}\left(
t\right) }{\partial t}$ of a related potential is used. The OfL potential is
transmuted to the effective OnL potential, and the gradient of the latter
can be calculated at the accessible value of $t$.

Equation (\ref{algoritmo2}) reminds others that have appeared in related but
different places and a few comments are in order. It is not totally
unrelated to those obtained in the cavity analysis of learning by Griniasty 
\cite{griniasty93}. The cavity and replica methods are not constructive,
they are used to determine the OfL performance of gradient descent learning
algorithms. The parameter ${\sf x}$ plays the role of the stiffness
parameter in the cavity analysis and that of $x=\lim_{\beta \rightarrow
\infty }\beta \left( 1-q\right) $ in the replica (symmetric) calculations.
With respect to the latter, Bouten {\it et al.} have, in their analysis of
OfL gradient descent learning, stressed the interpretation of replica
results in terms of cavity arguments.

But this effect of transmutation of potentials has been seen before in \cite
{kinouchi92}\cite{kinouchi96}. These works were done in the context of the
variational-optimization method. Its purpose is to determine a potential
that leads to maximum performance by functionally extremizing a performance
measure such as the generalization error with respect to the potential. For
some architectures it has been applied to both OnL and OfL learning in the
thermodynamic limit in order to determine maximum possible generalization.
It was found \cite{kinouchi96}, that for the single layer perceptron,
equation (\ref{effective}) gives precisely the relation between the optimal
generalization OnL and OfL potentials. The same relation holds in
unsupervised learning \cite{vdbroeck96}. Up to now this relation (eq.\ref
{effective}) seemed little more than accidental, but now can be seen as a
consequence of approximating OfL by the closest (in the sense of
Kullback-Leibler divergence) OnL learning scheme.

Equation \ref{deltac} describes the annealing of the tensorial learning
rate. Several works (e.g. \cite{amari67} ,\cite{barkai95}) have stressed the
need for an OnL learning rate annealing. The need comes from the fact that
once an estimate is close to a minimum of the potential, the step size
should be reduced in order not to overshoot. The analogous of an annealing
rate in an OfL problem appears e.g. in \cite{fontanari92}, where a
performance is improved by choosing a parameter of the potential (there, the
threshold $\kappa $ of a relaxation algorithm)\ from the knowledge of the
size of the learning set. This appears automatically in the variational
optimized potentials both OnL and OfL \cite{kinouchi93}\cite{kinouchi96}.
The origin of the need for annealing was thought to be the same. However,
here, as in the work of Opper, it can be seen that even if an OfL potential
is not annealed the imposition of minimal information loss will anneal the
OnL learning rate.

The case of the single layer perceptron with multiplicative noise, a
nonsmooth model, is interesting and we discuss it a little further. In \cite
{kinouchi96} the OfL potential that implements the Bayes bound for
generalization of Opper and Haussler was determined. If this potential is
used in equations (\ref{deltaw}) and (\ref{deltac}) the Bayes OnL algorithm
found by Solla and Winther is reobtained \cite{sollaw98}. They however could
not claim that their algorithm was the gaussian approximation to the OfL
Bayes because Opper's derivation (as theirs) is only valid for smooth
models. However, it is quite tempting to study the resulting equations for
nonsmooth models. From the point of view of designing learning algorithms
this is certainly acceptable. We have shown that they can actually claim
that the resulting algorithm , which they called Bayes OnL is the gaussian
approximation to an algorithm which indeed saturates the Bayes OfL\ limit.
For this model, the off-diagonal terms of the covariance tend to be smaller
than the diagonal by a factor of $\sqrt{N}.$ Asymptotically the covariance
tends to become diagonal and the asymptotic performance - as measured by the
generalization error - for $N\rightarrow \infty $ is the same as that of the
variational optimized algorithm.

To understand how the annealing is working, we analyze a smooth potential $V$
that is flat for large absolute values of the stability. For negative values
it saturates at a positive value, while for positive stabilities it goes to
zero. In the transition region it decays monotonically. This kind of
potential is quite sensible, actually the optimal one we discussed above is
of this type. The second derivative that enters the annealing equation is
positive if the example is correctly classified, and negative if not. This
means that the system is estimating on-line its performance. If in error, it
reacts by increasing the estimate of the variance of the posterior
distribution and in that manner, allowing larger corrections to be made to
the current estimate ${\bf \hat{w}}$. When an example is correctly
classified, then the system will start making smaller weight estimate
adjustments. Actually this is consistent with the idea, exposed e.g. \cite
{kinouchi93}\cite{caticha98}, that adaptive annealing schemes should depend
on the estimate of the generalization error.

>From an argument similar to Opper \cite{opper98}, the covariance annealing
is governed by 
\begin{equation}
\lim_{\mu \rightarrow \infty }\frac{C^{-1}}\mu =J_V\left( {\bf w}^{*}\right), 
\end{equation}
where the matrix $\left( J_V\left( {\bf w}^{*}\right) \right) _{ij}=%
\overline{\partial _i\partial _j{\cal E}_{{\sf x}}\left( t\right) }$ , and
the overbar indicates average over the examples distribution. This is not in
general Fisher's Information matrix, but it is expected to be so for some
cases. These include the additive noise case for the perceptron with the
optimal potential \cite{biehl95}, the unsupervised learning case \cite
{vdbroeck96} and the linear perceptron\cite{kinouchi95} , where the OnL
performance is asymptotically efficient. It is expected to differ in cases
such as the perceptron learning from a spherical distribution of examples in
the presence of multiplicative noise, since then OnL can achieve only twice
the error of the Bayes algorithm. It is possible that further studies of
this system of equations can shed light on this exact factor of 2.

Follwing Solla and Winther, we have not resisted the temptation to apply our
algorithms to potentials which do not satisfy the conditions of smoothness.
In particular an interesting case is the Perceptron algorithm of Rosenblatt
applied to a perceptron in a noiseless student-teacher scenario. The OfL
potential can be defined by $V_R\left( \lambda \right) =-\lambda \Theta
\left( -\lambda \right) $, where $\lambda =\sigma {\bf w.}S{\bf /}N.\;$A
possible prescription for the weights can be obtained by simulated
annealing. The interest resides in the fact that the generalization error
decays as $\alpha ^{-1}$OfL but only as $\alpha ^{-\frac 13}$ OnL. The
relevant quantity is the effective OnL energy ${\cal E}_{{\sf x}}\left(
t\right) $. The modulation function, $-\partial _t{\cal E}_{{\sf x}}\left(
t\right) \;$is 
\begin{equation}
\lim_{\beta \rightarrow \infty }-\partial _t\ln \int d\lambda \exp -[\beta
V_R(\lambda )+\frac{(\lambda -t)^2}{2{\sf \tilde{x}}}]=\frac 1{\sqrt{2\pi 
{\sf \tilde{x}}}}\frac{e^{-t^2/(2{\sf y})}}{H(\frac{-t}{\sqrt{{\sf \tilde{x}}%
}})}, 
\end{equation}
where ${\sf \tilde{x}}=S_iC_{ij}S_j/N\;$and $H(x)=\int_{-\infty }^x\exp
(-z^2/2)dz/\sqrt{2\pi }$. This is surprisingly close to the optimal OnL
modulation function. Even the annealing, which affects ${\sf y}$ is similar
and from equations (\ref{algoritmo}, \ref{annealing} )the OnL generalization
error decays as $\alpha ^{-1}$.

To conclude we have studied the first approximation OnL which is
(Kullback-Leibler) closest to potential learning OfL. Somewhat surprisingly
the OnL potential ${\cal E}_{{\sf x}}\left( t\right) $ is not the same as
the OfL $V(\lambda ).$The most striking feature is that they depend on
different quantities. The former on $t,$ the stability prior to learning,
and it could not be otherwise for the post presentation stability is
unknown. The latter, on the stability, which will tend, in equilibrium to
the OfL (equilibrium) post presentation stability. A second feature is
expected, the energy consists of a pure energy term $V$ associated to the
new term plus another that reflects the presence of previously presented
examples.

We refer to this as a first approximation since a systematic expansion can
be implemented \cite{evaldo00}. The infinite (formal) series shows that OfL
equilibrium is attained by parameters and hyperparameters updates that
involve only the effective OnL potential without making reference to the OfL
potential. In connection to this we look at the question \cite{engel00} of
what it means to learn OfL with a potential that is infinite for negative
stabilities. Gradient descent can only start if the current estimate is
within Version Space (VS)? This is the case of the noiseless perceptron
optimal potential mentioned above \cite{kinouchi96}. While this issue is not
totally closed, a tentative answer starts by noticing that the effective OnL
potential can be used even outside VS. A question immediately follows, and
it might be attacked in the future by the techniques of dynamical replicas 
\cite{mace98}. If the effective OnL potential is used iteratively in
learning from a restricted learning set, what will be the asymptotic time
state?

\ack

We acknowledge interesting discussions with M.Copelli, O.Kinouchi, P.
Riegler, R.Vicente, and also with participants of the Workshop on
Statistical Mechanics -Max Planck Institute -Dresden (1999) where an early
version of this work was presented. EAO was supported by a fellowship from
the Funda\c {c}\~{a}o de Apoio \`{a} Pesquisa do Estado de S\~{a}o Paulo
(FAPESP) and the research of NC was partially supported by the Conselho
Nacional de Desenvolvimento Cient\'{\i}fico e Tecnol\'{o}gico (CNPq)

\end{document}